\documentclass[12pt]{article}

\pdfoutput=1

\usepackage[top=80pt,bottom=85pt,left=85pt,right=85pt]{geometry}
\usepackage{amssymb}
\usepackage{amsmath}
\usepackage{setspace}
\usepackage{sectsty}
\usepackage{cite}
\usepackage{bbm}
\usepackage[utf8]{inputenc}
\usepackage[T1]{fontenc}
\usepackage{comment}
\usepackage[english]{babel}
\usepackage{afterpage}

\usepackage[usenames]{xcolor}
\usepackage{graphicx,subcaption}
\usepackage{setspace}
\usepackage{float}
\usepackage{booktabs}
\usepackage[debug,pageanchor=false]{hyperref}
\definecolor{link}{rgb}{.8,.15,.1}
\definecolor{pigment}{rgb}{0.36, 0.54, 0.66}
\definecolor{pigment2}{rgb}{0.19, 0.55, 0.91}
\definecolor{pigment3}{rgb}{0.2, 0.2, 0.6}
\definecolor{light-gray}{gray}{0.75}
\hypersetup{colorlinks=true,linkcolor=link,citecolor=link,urlcolor=link,linktocpage}

\usepackage{array,multirow,booktabs,longtable}
\usepackage{mathrsfs}
\usepackage{tikz-cd} 
\usetikzlibrary{backgrounds, arrows,calc,shapes,decorations.pathreplacing, decorations.markings, automata,positioning}
\tikzset{%
  >={Latex[width=2mm,length=2mm]},
            base/.style = {rectangle, rounded corners, draw=black,
                           minimum width=4cm, minimum height=1cm,
                           text centered, font=\sffamily},
  activityStarts/.style = {base, fill=orange!15},
       startstop/.style = {base, fill=orange!15},
    activityRuns/.style = {base, fill=orange!15},
         process/.style = {base, minimum width=2.5cm, fill=orange!15,
                           font=\ttfamily},
}

\newcommand{\red}[1]{}

\tikzset{
        cvertex/.style={circle,draw=black,inner sep=1pt,outer sep=3pt},
        vertex/.style={circle,fill=black,inner sep=1pt,outer sep=3pt},
        star/.style={circle,fill=yellow,inner sep=0.75pt,outer sep=0.75pt},
        tvertex/.style={inner sep=1pt,font=\scriptsize},
        gap/.style={inner sep=0.5pt,fill=white}}

\tikzstyle{mybox} = [draw=black, fill=blue!10, very thick,
    rectangle, rounded corners, inner sep=10pt, inner ysep=20pt]
\tikzstyle{boxtitle} =[fill=blue!50, text=white,rectangle,rounded corners]

\newcommand{\cc}{\mathbb{C}}
\newcommand{\zz}{\mathbb{Z}}
\newcommand{\pp}{\mathbb{P}^1}

\newcommand{\todo}[1]{}
\renewcommand{\todo}[1]{{\color{red} TODO: {#1}}}
\renewcommand{\red}[1]{{\color{red} {#1}}}

\newcommand{\be}{\begin{equation}}  
\newcommand{\ee}{\end{equation}}  
\newcommand{\bea}{\begin{align}}
\newcommand{\eea}{\end{align}}
\newcommand{\bp}{\begin{bmatrix*}[r]}  
\newcommand{\ep}{\end{bmatrix*}}  
\newcommand{\bpp}{\begin{bmatrix}}  
\newcommand{\epp}{\end{bmatrix}}  
\newcommand{\bcd}{\begin{center}
\begin{tikzcd}}
\newcommand{\ecd}{\end{tikzcd} \end{center}}

\makeatletter
\@addtoreset{equation}{section}
\makeatother

\begin{document}


\begin{titlepage}

\begin{center}

\vskip .3in \noindent

{\Large \bf{A string theory realization of special unitary quivers in 3 dimensions}}

\bigskip\bigskip

Andr\'es Collinucci$^a$ and Roberto Valandro$^b$ \\

\bigskip


\bigskip
{\footnotesize
 \it

$^a$ Service de Physique Th\'eorique et Math\'ematique, Universit\'e Libre de Bruxelles and \\ International Solvay Institutes, Campus Plaine C.P.~231, B-1050 Bruxelles, Belgium\\
\vspace{.25cm}
$^b$ Dipartimento di Fisica, Universit\`a di Trieste, Strada Costiera 11, I-34151 Trieste, Italy \\
and INFN, Sezione di Trieste, Via Valerio 2, I-34127 Trieste, Italy	
}

\vskip .5cm
{\scriptsize \tt collinucci dot phys at gmail dot com \hspace{1cm} roberto dot valandro at ts dot infn dot it}

\vskip 1cm
     	{\bf Abstract }
\vskip .1in
\end{center}
We propose a string theory realization of three-dimensional $\mathcal{N}=4$ quiver gauge theories with special unitary gauge groups. This is most easily understood in type IIA string theory with D4-branes wrapped on holomorphic curves in local K3's, by invoking the St\"uckelberg mechanism.
From the type IIB perspective, this is understood as simply compactifying the familiar Hanany-Witten (HW) constructions on a $T^3$. 
The mirror symmetry duals are easily derived. We illustrate this with various examples of mirror pairs.
\noindent

\vfill
\eject

\end{titlepage}

\tableofcontents

\section{Introduction} 
\label{sec:intro}
Hanany and Witten \cite{Hanany:1996ie} showed us how to build quiver gauge theories in three dimensions with eight supercharges, and unitary gauge groups. The setting was type IIB string theory, with D3-branes suspended between NS5-branes, D5-branes, and various arrangements thereof. Because type IIB theory enjoys S-duality, this allowed them to easily derive for each theory they considered, an IR dual theory related to it via \emph{3d mirror symmetry}, as conceived in \cite{Intriligator:1996ex}.
Alternative string theory setups were considered in the works \cite{deBoer:1996mp,Porrati:1996xi}.

It is then natural to wonder, what the mirror duals of quivers with \emph{special unitary} gauge groups are.
In order to render a unitary gauge group special unitary, Witten devised a field theory method in \cite{Witten:2003ya}, whereby he essentially gauges the topological $U(1)$ global symmetry present. The coupling then acts like a St\"uckelberg mass term for the original photon. In \cite{Hanany:1996ie}, Hanany and Witten apply this method to various examples. This requires them to introduce an auxiliary hypermultiplet with an appropriate coupling to the photon. Through this method, they are able to derive various mirror pairs. More elaborate examples of this can be found in \cite{Dey:2014tka}, and more generally even, in \cite{Dey:2020hfe}.

However, there is a vexing issue that persists: To our knowledge, no systematic direct method has been concocted to generate special unitary quiver in string theory\footnote{See \cite{Klebanov:2010tj,Benishti:2010jn} for a holographic method for gauging some  topological $U(1)$ symmetries in M-theory setups.}. This would be useful for several reasons: It would essentially prove the various mirror pairs that have been conjectured, provided we trust that S-duality is an exact symmetry of type IIB string theory; it might open the way for computing quantities directly via string theory methods, as opposed to relying only on field theory. Finally, this will open up a new avenue of investigation for so-called `bad theories', as defined by \cite{Gaiotto:2008ak}. Theories with too few flavors are problematic, because they have no known conventional QFT mirror duals. There has been research in this direction \cite{Assel:2017jgo}.

In this paper, we solve this problem by giving two explicit string theory constructions of special unitary quivers: One in type IIA and one in type IIB string theory. Our IIA construction will allow us to immediately see special unitary groups appear via a St\"uckelberg mechanism that involves the anomalous coupling of open and closed string modes on the D4-brane worldvolume. The IIB construction is related to it via fiberwise T-duality. In IIB, S-duality will allow us to readily derive mirror duals.
Moreover, 
the framework established in this paper will allow one to derive string theory setups for  unconventional mirror duals involving `bad theories'. We will show this for the case of $SU(2)$ with $2$ flavors.

\section{The St\"uckelberg mechanism in string theory}\label{Sec:Stuck}

We review the mechanism by which the worldvolume photon of a D-brane can gain a St\"uckelberg mass through its anomalous coupling to RR fields. The mechanism is well-known, and discussed in various paper such as \cite{Hanany:1997gh,Brunner:1997gf,Grimm:2011tb}. The authors have recently analyzed this in an analogous setup for five-dimensional field theories in \cite{Collinucci:2020jqd}. We will closely follow that approach here.

A single Dp-brane in $10$d spacetime gives rise to the combined system:
\be
S_{\rm tot} = S_{{\rm DBI}} + S_{{\rm WZ}} + S_{{\rm SUGRA}}\,.
\ee 
Let us focus on the subsector of supergravity involving the RR $(p-1)$-form potential $C_{p-1}$, and the open/closed string anomalous coupling of the gauge field $A$ with $C_{p-1}$. The relevant action is\footnote{We have fixed $\kappa_{10}=1/2$ for simplicity.}
\be
S[A, C_{p-1}]  = \int_{\mathbb{R}^{10}} d C_{p-1} \wedge \star d C_{p-1}  + \mu_p \int_{\mathbb{R}^{p+1}} F \wedge C_{p-1}  \:.
\ee

Suppose now that the transverse directions to the brane are compactified\footnote{Ultimately, we will assume a more subtle situation.}  on a torus $T^{9-p}$ of volume $V$ so that now we have
\be\label{EffActionStu}
S_{eff}[A, C_{p-1}]  = \int_{\mathbb{R}^{p+1}}{\Big (}  d C_{p-1} \wedge \star d C_{p-1} + \mu'_p  F \wedge C_{p-1}  {\Big )}\:.
\ee
where  $\mu_p'\equiv \frac{\mu_p}{V^{1/2}}$ and where we have normalized the dimensionally reduced fields $C_{p-1}$.

The $(p-1)$-form can be dualized to a scalar propagating in $d+1$ dimensions. Let us review the dualization procedure: we start by the following action depending on the field $A$,  a $p$-form $T_p$ and a scalar $\sigma$:
\begin{align}
& \tilde S[A,  T_p, \sigma]  = 
  \int_{\mathbb{R}^{p+1}}\Big ( \,T_p \wedge \star T_p + \mu'_p A \wedge T_p  +  \sigma \, d T_p \,\Big ) \:.
\end{align}
Here, we are working in Lorentzian signature, so the path integral has a factor of the form $\exp(i \int \sigma d T_p)$ in the integrand. Integrating by parts and using Stokes' theorem, we see that, if we shift $\sigma \mapsto \sigma + c$, we get a phase of the form
\be
\exp(i c \int_{S^p} T_p) = \exp(i c n)\,,
\ee
since $T_p$ has quantized flux in the full string theory. This emplies that the scalar $\sigma$ is circle-valued.

Now we can proceed in two ways:
\begin{enumerate}
\item We can integrate out the scalar $\sigma$: this forces $dT_p=0$;  in a topologically trivial worldvolume, this implies it is exact, i.e. $T_p = d C_{p-1}$. We then come back to the original action \eqref{EffActionStu}, after a simple integration by parts.
\item We can integrate out $T_p$: we are then left with 
\be\label{EffActionStuSigma}
S_{eff}[A, \sigma] = \tfrac14 \int d^{p+1} x | d \sigma-\mu'_p A |^2\:.
\ee
\end{enumerate}

The action \eqref{EffActionStuSigma} is equivalent to \eqref{EffActionStu} and is a St\"uckelberg mass term for $A$. Hence, when the coupling $\int F\wedge C_{p-1}$ is present, the involved gauge field becomes massive and only the global part of the $U(1)$ remains in the low energy theory.

Note, that it is crucial for the RR field to be dynamical. This is why $N$ D-branes in flat spacetime carry a $U(N)$ gauge group. In order for the overall $U(1)$ to gain a St\"uckelberg mass, there must be an effective dimensional reduction of the RR-field to the worldvolume of the brane. As we have seen, this can be achieved  by fully compactifying the transverse space. 
Another option, however, is to have a non-compact spacetime that admits a harmonic $k$-form $\omega_{k}$ that is normalizable in the $p-9$ dimensions that are transverse to the brane worldvolume, such that if we take the Ansatz
\be
C_{p-1} = \omega_{k} \wedge c \:,
\ee
with $c$ a $(p-1-k)$-form propagating in $(p+1)$-dimensions,
then the kinetic term of the RR-potential will effectively localize around the brane locus:
\be
\int d^{10} x \, |d C_{p-1}|^2 = \int d^{10} x |\omega_{k}|^2  |d c|^2 \sim  \int d^{p+1}x |dc|^2\:.
\ee

Throughout this paper, we will rely on a hybrid situation: Our target space will be $\mathbb{R}^3 \times \widehat{\mathbb{C}^2/\Gamma} \times T^3$, where the orbifold will be resolved, and a metric will be chosen such that normalizable 2-forms exist.

In particular, we will look at D4-branes wrapped on exceptional $\mathbb{P}^1$ inside the local K3. Hence, their transverse space will consist of the $T^3$ times (roughly) their normal bundle inside the K3, which will be non-compact, but nevertheless still admitting normalizable harmonic two-forms. We will therefore make the case that the coupling~$F \wedge C_3$ will make the $U(1)$ gauge field massive, so long as there is a $T^3$. Decompactifying this $T^3$, however, will make the $C_3$ field non-dynamical, thereby restoring the $U(1)$ gauge field.

\section{Strategy}
The goal of this paper is twofold: To show how to both build \emph{special} unitary 3d quivers \emph{and} derive their mirrors in string theory.

The first goal will be achieved in two ways: First we will show how the St\"uckelberg mechanism described in the previous section can be readily applied to type IIA string theory on local K3's times $T^3$, with D4-branes wrapping compact and non-compact holomorphic curves of said K3. The $T^3$ factor renders the appropriate bulk fields dynamical, thereby triggering one St\"uckelberg mechanism per gauge node, stripping it of its overall $U(1)$ factor.

Secondly, we will T-dualize this to the more familiar Hanany-Witten (HW) scenarios with D3, D5 and NS5-branes \cite{Hanany:1996ie}. We will then explain directly from the IIB perspective why having a $T^3$ factor eliminates $U(1)$'s for each gauge node.

The advantage of the IIA perspective is that the St\"uckelberg mechanism is easier to understand from a string theory perspective. The IIB explanation will require a field theory argument. However, the IIB viewpoint makes mirror symmetry easier to implement.

Then, we will be ready for our second goal: By applying S-duality in IIB, we will derive mirror symmetry for such quivers. In our final section, we will display various classes of examples to demonstrate the validity of this approach.

\section{IIA perspective} \label{sec:iiaperspective}

We start considering type IIA string theory with target space $\mathbb{R}^{1,2} \times T^3 \times \widehat{\mathbb{C}^2/\Gamma}$, whereby the last factor is a resolved orbifold of $\mathbb{C}^2$ by a discrete subgroup $\Gamma \subset SU(2)$. The metric (whether ALE or ALF) will admit as many (or more) normalizable two-forms $\eta_i$ as there are exceptional curves $C_i\cong \mathbb{P}^1$; here $i=1,...,r$ where $r$ is the rank of the ADE algebra associated with the discrete group $\Gamma$.
We choose local coordinates $x_0,...,x_9$, where the coordinates $x_{0,1,2}$ are along $\mathbb{R}^{1,2}$, $x_{3,4,5,6}$ are local coordinates along the resolved orbifold and $x_{7,8,9}$ are along the 3-torus $T^3$.

The type IIA string theory can be reduced to 3d. Let us do it in two steps for convenience: we first reduce on the resolved orbifold and then we compactify on $T^3$. Type IIA on $\widehat{\mathbb{C}^2/\Gamma}$ gives a 6d theory. The closed string fields are expanded along the normalizable two-forms $\eta_i$:
\begin{equation}
C_3 = \mathcal{A}^i_M dx^M \wedge \eta_i + ... \qquad\qquad  B_2=b^i\eta_i + ...
\end{equation}
with $M=0,1,2,7,8,9$. Since $\eta_i$ have compact support on $\widehat{\mathbb{C}^2/\Gamma}$, the 6d vectors $\mathcal{A}^i_M$ and scalars $b^i$ propagate in 6d. Expanding the hyperK\"ahler two-forms $\omega_k$ ($k=1,2,3$) of the local K3 along the same normalizable two-forms, one obtains the metric moduli; they propagate in 6d as well. Fixing a K\"ahler structure, one can take the K\"ahler form to be $J=\omega_3$ and the holomorphic (2,0)-form $\Omega=\omega_1+i\,\omega_2$. The metric moduli can be read off from
\begin{equation}
J = \xi^i \eta_i + ... \qquad\qquad  \Omega=\zeta^i\eta_i + ...
\end{equation}
where $\xi^i$ are 6d real scalars and $\zeta^i$ are 6d complex scalars.
The vector $A^i_M$ and the scalars $\xi^i,\zeta^i,b^i$ sit together in a 6d $\mathcal{N}=2$ vector multiplet. 
We now compactify the 6d theory on $T^3$. The 6d  vector $\mathcal{A}^i_M$ gives rise to a 3d vector $\mathcal{A}^i_\mu$ and three 3d scalars $\mathcal{A}^i_{7,8,9}$. All these fields sit together with the 3d scalars $\xi^i,\zeta^i,b^i$ into a 3d $\mathcal{N}=8$ vector multiplet.

We now introduce D4-branes. 
These are graphically summarized in Table~\ref{fig:IIArealization}.
%

\begin{table}[h!]
\be
\begin{tabular}{c|cccccccccc}
 & 0 & 1 & 2 & 3 & 4 & 5 & 6 & 7 & 8 & 9\\
D4$_G$ & --- & --- & --- & --- & ---  &   &   & &  &  \\
D4$_F$ & --- & --- & --- &  &   &   --- &  ---  &   & &  \\ 
K3 &  &  &  &---  &   ---   &    --- & ---& & \\ 
 \end{tabular}\nonumber
 \ee
\caption{IIA realization. The gauge branes wrap exceptional curves (along $x_3,x_4$), and the flavor branes wrap copies of $\mathbb{C}$ (along $x_5,x_6$).}
    \label{fig:IIArealization}
\end{table}

The gauge D4$_G$-branes  wrap the compact curves $C_i$; the flavor D4$_F$-branes wrap $\mathbb{C}$'s transverse to $C_i$. The D4-branes are BPS in the given target space if the wrapped loci are holomorphic with (the same) one choice of the K\"ahler structure on the local K3.
These BPS D4-branes break half of the supersymmetries. Correspondingly the bulk vector multiplets break into $\mathcal{N}=4$ vector multiplets $(\mathcal{A}_\mu^i,b^i,\zeta^i)$ and  $\mathcal{N}=4$ hypermultiplets $(\xi^i,\mathcal{A}_7^i,\mathcal{A}_8^i,\mathcal{A}_9^i)$.

Let us first consider the case in which $T^3$ decompactifies to $\mathbb{R}^3$.
 Wrapping $N_i$ D4-branes on $C_i$ gives rise, at low energies, to an $\mathcal{N}=4$ $U(N_i)$ vector multiplet, with no adjoint hypers, owing to the fact that such a curve has a negative normal bundle $\mathcal{N}_i \cong \mathcal{O}(-2)$, i.e. is `rigid'.

When two such curves $C_i$ and $C_j$ intersect, bifundamental strings stretched between them will give rise to a hypermultiplet in the $(\bar{N_i}, N_j)$. 

We also consider non-compact D4-branes stretching along copies of $\mathbb{C}$ that intersect the compact curves at points. This will give rise to hypers in the fundamental of the intersected curves. Let us label such non-compact curves $D^\ell$.

This class of constructions allows us to consider all quiver gauge theories shaped like Dynkin diagrams.
As we said, each $C_i$ gives rise to a round node with a $U(N_i)$ gauge group. Round nodes intersect according to the Dynkin graph whose corresponding Cartan matrix is determined by the intersection matrix $A_{i j} = -C_i \cdot C_j$. In addition, putting $N_f^\ell$ D4-branes on $D^\ell$, contributes a square flavor node with flavor group $SU(N_f^\ell)$, attached to the corresponding compact $C_j$ according to the intersection $C_j \cdot D^\ell$.

All such quivers have unitary gauge groups. However, if we now compactify on $T^3$ the non-compact $x_{7, 8, 9}$ directions that are normal to the branes and to the resolved local K3, the St\"uckelberg mechanism described in Section~\ref{Sec:Stuck} will eliminate all $U(1)$ factors, leaving us with special unitary gauge groups at each node. In particular, one has the following coupling on the D4$_G$-branes:
\be
\int_{D4_G^i}  F^i \wedge C_3
\ee
where $F^i$ is the field strength of the diagonal U(1) on the D4 stack wrapping $C_i$.
This coupling  induces  mixed Chern-Simons terms in 3d
\be
\sum_i \int_{\mathbb{R}^{1,2}}   F^i\wedge \mathcal{A}^i_\mu dx^\mu \:.
\ee
This makes the $U(1)$ gauge fields massive. 
In fact, at the 3d level, what we have done corresponds to gauging the topological $U(1)_T$ symmetries. This term is reminiscent of the BF-terms considered in \cite{Kapustin:1999ha}, whereby one subsequently renders the background fields dynamical.

This happens so long as there is a $T^3$.
Decompactifying this $T^3$, however, will make the $C_3$ field non-dynamical (ungauging the corresponding topological symmetry), thereby restoring the $U(1)$ gauge field.

Let us address the consequences of compactifying on $T^3$ for the moduli space of the effective theory. The full moduli space of the D4-branes now becomes a compact space (i.e. an $S^1$-fibration over $T^3$). However, the low-energy effective theory on the D4-branes only sees an open patch of the base $T^3$ that is diffeomorphic to a 3-disk. This is understood as follows: In order to flow to the infrared, we set a cutoff scale (something below the KK scale set by the radii of the $T^3$). This imposes an upper cutoff in the vev's that the worldvolume adjoint scalars can take, hence bounding the D4-branes' movements to an open disk. 
In this way, we recover a Coulomb branch of the form of an $S^1$-fibration over $\mathbb{R}^3$, which is fully compatible with a hyper-K\"ahler structure.

As far as the R-symmetry is concerned, there is an analogous paradox with a resolution. In the target space, the $T^3$ no longer enjoys an $SO(3)$-isometry. This would seem to spoil the expected $SO(3)_R$-subgroup of the R-symmetry group acting on the Coulomb branch. However, the 3 adjoint scalars are actually sections of the normal bundle to the D4-branes, which still admits an $SO(3)$-action, since each fiber is diffeomorphic to $\mathbb{R}^3$. Hence, the effective field theory behaves as expected for an $\mathcal{N}=4$ gauge theory.

Note, that the St\"uckelberg mechanism survives this IR limit, since it relies on KK-zero modes of bulk fields reduced along the $T^3$. Those modes remain massless all the way down the RG flow, and continue to ensure the mass term for the photon.

\section{IIB perspective}
\subsection{Special unitary linear quivers}
Having established how to construct special unitary Dynkin quivers in type IIA string theory, we restrict our attention to $A$-type K3-surfaces,\footnote{I.e. resolutions of $A_n$ orbifold singularities.} which admit a multi-Taub-NUT metric. This means that these spaces are $S^1$-fibered, allowing for a T-duality, albeit at the cost of generating NS5-branes whenever the $S^1$-fiber collapses. This lands us in the familiar territory of Hanany-Witten setups.  

Symbolically, we will say that we T-dualize the $x_4$ coordinate (see Table~\ref{fig:IIArealization}). This paints for the situation shown in Table~\ref{fig:IIBrealization}.
\begin{table}[h!]
\be
\begin{tabular}{c|cccccccccc}
 & 0 & 1 & 2 & 3 & 4 & 5 & 6 & 7 & 8 & 9\\
D3$_G$ & --- & --- & --- & [---] &   &   &   & &  &  \\
D5$_F$ & --- & --- & --- &  &  --- &   --- &  ---  &   & &  \\ 
NS5 & --- & --- & --- &  &      &     & &--- & ---&---
 \end{tabular}\nonumber
 \ee
\caption{T-dual IIB realization. The result is a Hanany-Witten scenario after we decompactify the $x_4$ direction.}
    \label{fig:IIBrealization}
\end{table}

\begin{figure}[t!]
\centering
\includegraphics[width=1.0\textwidth]{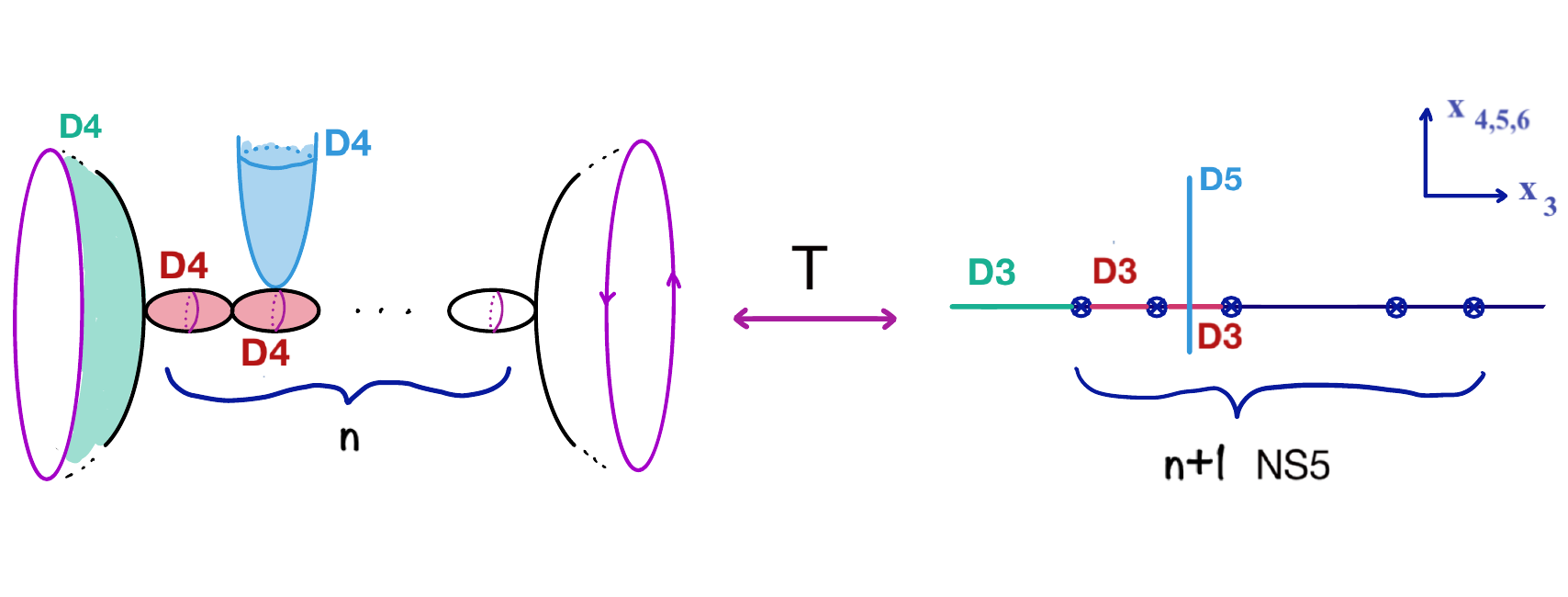}
\caption{T-duality between IIA with D4 branes on a local K3 (we have drawn the $A_n$ case with $n$ exceptional spheres) and IIB with NS5/D5 branes and D3 branes.}
\label{fig-setupIIAIIB}
\end{figure}

Wherever we had a $\pp$ in IIA, we now have a pair of NS5-branes separated by an interval in the $x_3$-direction. D4-branes on such $\pp$'s become D3-segments suspended between said NS5's (see Figure~\ref{fig-setupIIAIIB}). 

For the non-compact D4's, two things can happen: If the D4 intersected the $\pp$ at a generic point, it gets dualized to a D5 intersecting the D3-segment. In Figure~\ref{fig-setupIIAIIB}, this is depicted by the blue D4 and D5-branes. The other possibility is that the non-compact D4 extends along a holomorphic curve that contains the Taub-NUT circle. In this case, it will turn into a semi-infinite D3. In Figure~\ref{fig-setupIIAIIB}, this is depicted by the green D4 and  D3-branes. Note, that a `green' D4-brane can freely move to a more generic curve and become a `blue' brane. After T-duality, this corresponds to performing a Hanany-Witten transition.

It is interesting to see how a simple motion of the non-compact D4 in IIA gets turned into such a drastic move in IIB: It corresponds to moving a D5 that intersects a D3-segment through one of the NS5's (the classic HW move), and then sending the D5 off to infinity, leaving us with a semi-infinite anomalously created D3. 

Having performed this T-duality, we now have a Hanany-Witten setup, but with one direction of the D5's on a circle. Taking this circle to infinite size will not affect the quiver gauge theories. At worst, it will generate a KK tower in 10 spacetime dimensions, but nothing that will be dynamical in 3d. Hence, we freely decompactify it. The resulting setup is drawn in Figure~\ref{fig-setupIIAIIB}

In order to transform $U(N)$ gauge groups into $SU(N)$ gauge groups in the IIA setup, we needed to compactify $x_{7, 8, 9}$ on a $T^3$.

{\bf Therefore, our claim is that one can obtain special unitary quivers from Hanany-Witten scenarios compactifying the NS5-directions that are transverse to the D3-branes on a $T^3$.} 

As explained in Section \ref{sec:iiaperspective}, despite the compactification on $T^3$, the moduli space, as seen by the low energy worldvolume theory on the D3-branes, is still non-compact. Similarly, its R-symmetry remains an $SO(3) \times SO(3)$.

Let us now explain this phenomenon directly in type IIB string theory. From the field theory perspective, one way to turn $U(N)$ gauge groups into $SU(N)$ gauge groups is to gauge the topological $U(1)_T$. Define the topological $U(1)_T$ current of a gauge node as
\be
J_T = \star d A\,,
\ee
where $A$ is the $U(1) \subset U(N)$, and $\gamma$ the dual photon. Then we introduce a background photon $A_T$, and couple it to $J_T$ via a term of the form
\be
J_T \wedge \star A_T \sim F \wedge A_T\,.
\ee
If $A_T$ becomes dynamical, then the $U(1)_T$ is gauged, and at the same time the original gauge $U(1)$ acquires a St\"uckelberg mass. The newly gauged $U(1)_T$, however, does not confer electric charge to the hypers, hence the quiver is indeed altered into a special unitary quiver.
How is this term realized in IIB string theory? 
A clear candidate is the anomalous D3-worldvolume coupling to the RR 2-form
\be
\int_{\rm D3} F \wedge C_2\,.
\ee
One might now object, that $C_2$ will not become dynamical in 3d simply by putting $x_{7, 8, 9}$ on a $T^3$. However, the presence of NS5-branes implies there are normalizable modes trapped on their worldvolume, which does get reduced to 3d. We can see this by tracking the T-duality process: In IIA on the multi-Taub-NUT background TN$_N$, there was a bulk 6d $U(1)^{N}$ symmetry provided by choosing the Ansatz for the RR 3-form 
\be
C_3 = \omega_2^i \wedge A_i\,,
\ee
where the $\omega_2^i$ are the $N$ normalizable 2-forms. This symmetry gets enhanced to $A_{N-1}$ by wrapping D2-branes on the compact spheres.\footnote{Actually it enhances to $U(N)$ in the ALF case.}

Our fiberwise T-duality will turn these $C_3$ modes into $C_2$ modes whose energy densities localize around the NS5-branes.\footnote{The D2-branes will turn into D1-strings stretched between the NS5-branes.} Therefore, by compactifying the NS5-branes to 3d, the anomalous coupling will do precisely what we expect: To gauge the topological $U(1)_T$ for each gauge node.

\subsection{Mirror symmetry}
Having established that $SU(N)$ linear quivers can be easily created in IIB Hanany-Witten setups by simply compactifying the three NS5 worldvolume directions that are transverse to the D3-segments, let us see what this means for mirror symmetry. As in the original HW paper, this duality is achieved by performing an S-duality. NS5-branes and D5-branes are exchanged, and the D3-segments remain as they are. 

However, after the S-duality, it is the D5-branes that will see three of their directions compactified. What are the field theory consequences of this? Assuming that we have kept the D5-branes apart, the consequences are simple, the Cartan subgroup of the Higgs branch flavor symmetry is gauged. 

\begin{figure}[t!]
\centering
\includegraphics[width=0.75\textwidth]{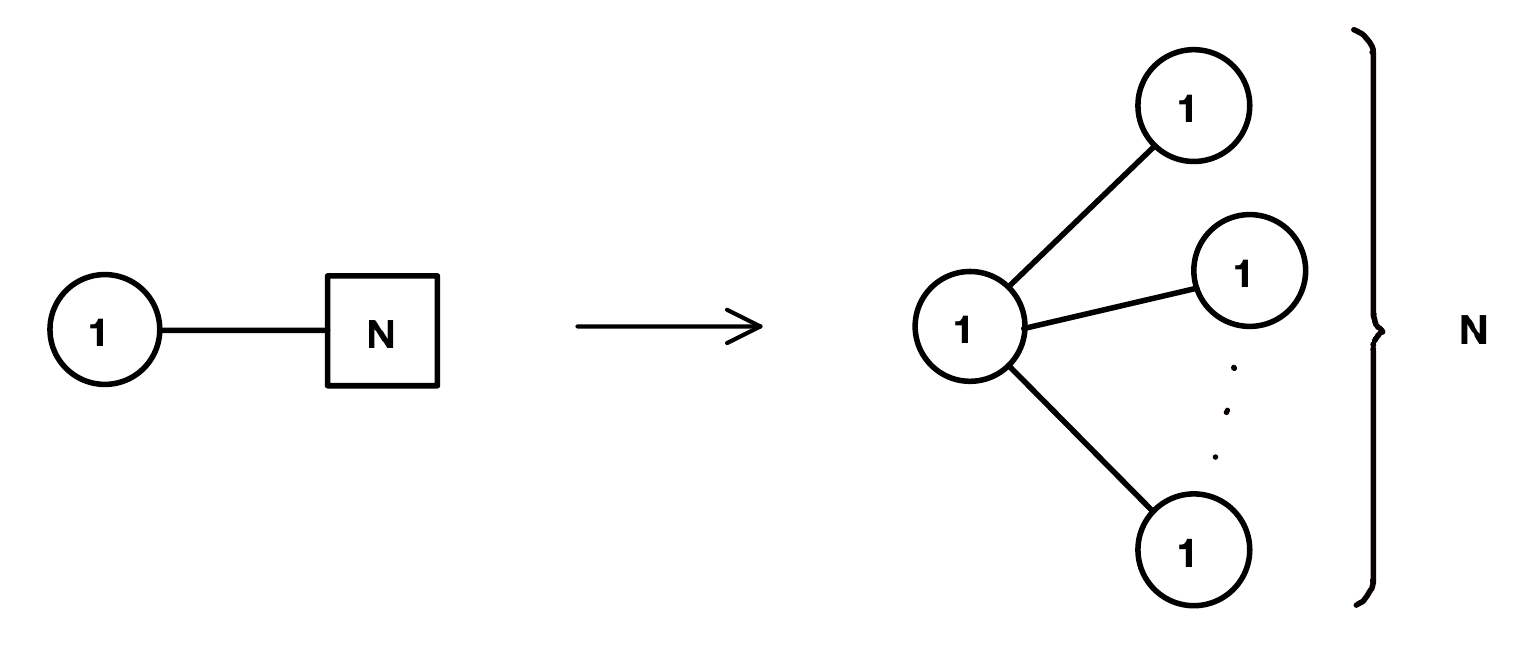}
\caption{The quiver after compactifying the directions transverse to the D5's.}
\label{fig-SUNvsNU1}
\end{figure}

Let us unpack this statement: Assuming we arrange for the D5-branes to be within the suspended D3-segments (as opposed to being outside on the extremities), then the hypermultiplets come from D3-D5 fundamental strings. For $N_f$ such D3-D5 intersections, the flavor symmetry is $SU(N_f)$. This is true regardless of the relative positions of the D5-branes. If one were to gauge this flavor group with a background gauge field, necessarily, one would use the gauge field on the worldvolume of the D5-branes. In the case of separated D5-branes, there is a $U(1)^{N_f}$ 6d gauge symmetry. By putting this on a $T^3$, $N_f$ photons become dynamical in 3d. 

The end results is that any `square node' in the quiver with $N_f$ will `implode' into $N_f$ distinct $U(1)$ gauge nodes, all attached to the same original gauge node. See Figure~\ref{fig-SUNvsNU1}. A round node with the number $n$ inside means a $U(n)$ gauge group. We will use this notation in the rest of the paper.\footnote{When the gauge group is $SU(n)$, we will explicitly write it in the quiver.}

This is precisely what was expected from field theory arguments. The topological $U(1)$'s of the original theory get mapped under mirror symmetry to the Cartan subgroup of the flavor group acting on the Higgs branch. In the next section, we demonstrate all of this with various examples.

\section{Examples}
We will present several families of examples of mirror pairs in this section. We will always refer to the starting theory (with special unitary gauge groups) as the `A-theory', and the mirror theory as the `B-theory'. The B-theory will always have unitary gauge nodes.

\subsection{Enhancement of the Higgs branch}
Before jumping into mirror symmetry, we would like to explore one important phenomenon. After compactifying the directions $x_{7,8,9}$ longitudinal to the NS5-branes (but transverse to the D3-segments), $U(N)$ gauge groups get reduced to $SU(N)$'s. Since a vector multiplet becomes unavailable to participate in the Higgs mechanism, the expected quaternionic dimension of the Higgs branch should increase by one. In fully Higgsable models, this is manifest in the formula:
\be
{\rm dim}_{\mathbb{H}} \mathcal{H} = \# {\rm hypers} - \# {\rm vectors}\,,
\ee
but can be true for other cases too.
More precisely even, when transitioning from $U(N)$ to $SU(N)$ we expect to see baryonic branches emerge.

\begin{figure}[ht!]
\centering
\includegraphics[scale=.5]{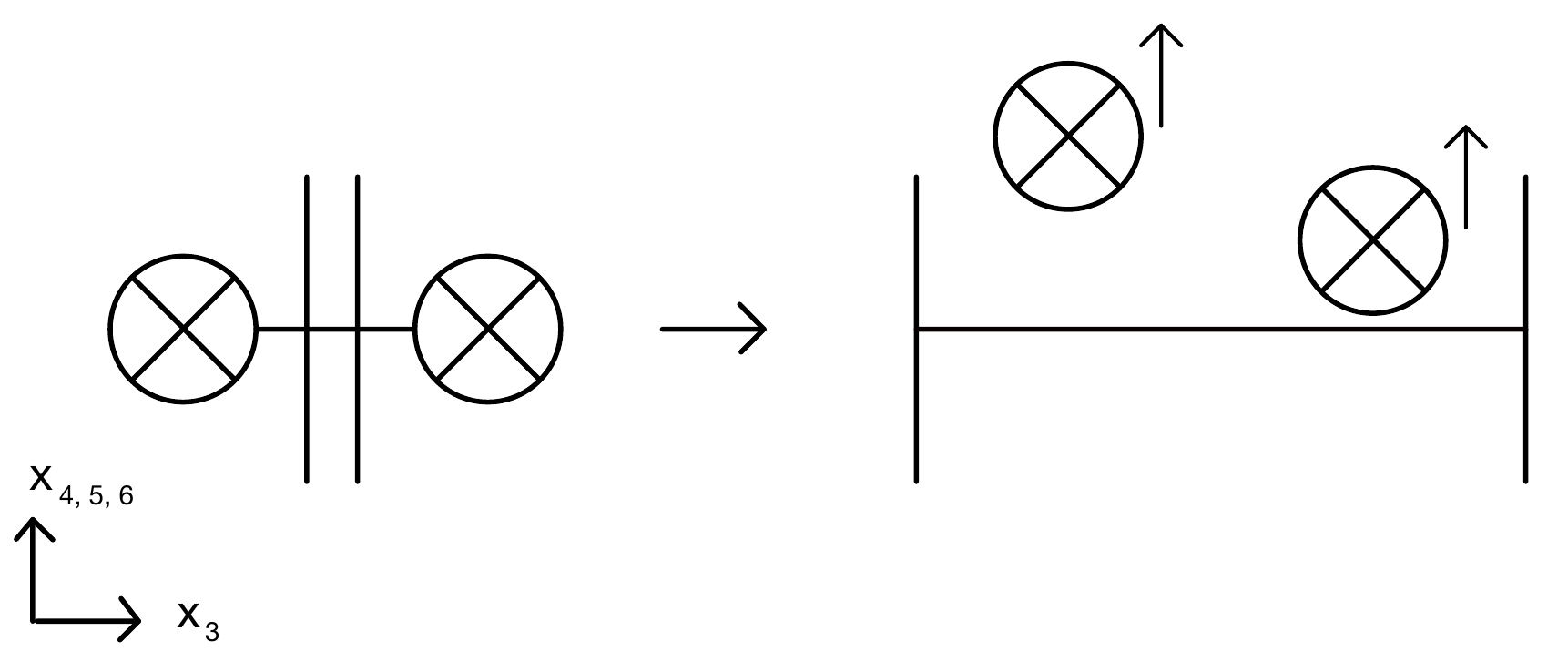}
\caption{$SU(1)$ plus two `charged' hypers.}
\label{fig:hw_sun_nf_jump}
\end{figure}

Can we see this in our string theory realization? The answer is yes. The NS5-worldvolumes are now three-dimensional, hence their relative positions correspond to dynamical fields in 3d. 
In order to see the jump up by one we illustrate the simple case of `$SU(1)$' with two flavors in Figure~\ref{fig:hw_sun_nf_jump}. The transition from $U(1)$ to `$SU(1)$' should make a one-dimensional Higgs branch grow to a two-dimensional one. The original dimension is accounted for by the motion of the D3-segment along the two external D5-branes. In addition, we see that two NS5-branes are now free to move in the $x_{4,5,6}$ directions. However, only their relative positions are physical, since their center of mass can be reabsorbed by moving the D3-segment. 
Together with the relative dual photon, this makes one linear multiplet with which the St\"uckelberg coupling is achieved, thereby giving mass to one vector multiplet, and freeing up one Higgs branch direction. Hence, in order to count extra Higgs branch quaternionic dimensions, we simply need to count the number of mobile NS5-branes minus one.

\subsection{SU(N) with $N_f$ flavors for $N_f \geq2 N$}
\begin{figure}[ht!]
\centering
\begin{subfigure}{.4\textwidth}
  \centering
  \includegraphics[width=1.0\linewidth]{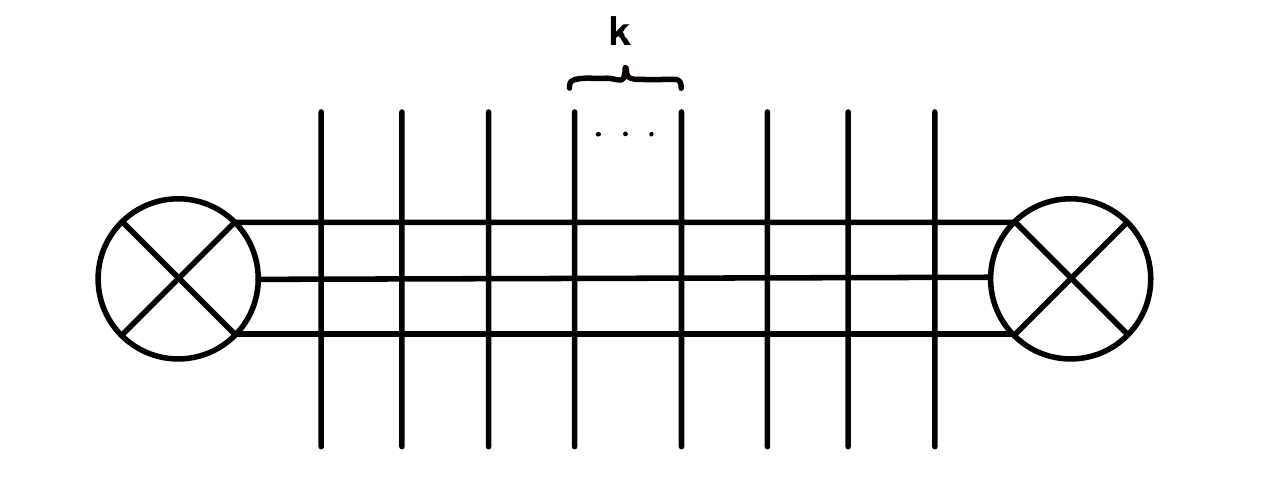}
\end{subfigure}$\,\,\cong\,\,$%
\begin{subfigure}{.5\textwidth}
  \centering
  \includegraphics[width=1.0\linewidth]{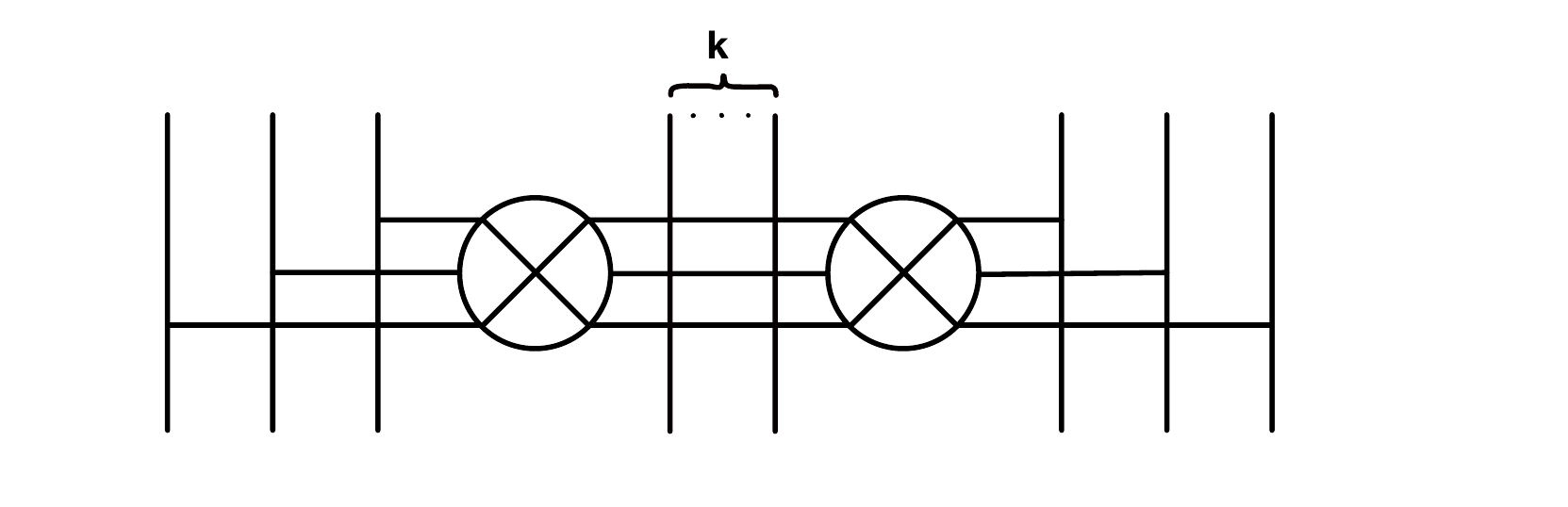}
\end{subfigure}
\caption{HW equivalent configurations for $SU(3)$ with $N_f=6+k$ flavors.}
\label{fig-su3_6}
\end{figure}

Let us now switch to the study of mirror symmetry for `good' quivers. This ensures full Higgsability, and the existence of a mirror dual that admits a conventional Lagrangian description in the UV. In Figure~\ref{fig-su3_6}, we present the case of $SU(3)$ with $N_f=6+k$ flavors as our A-theory.

\begin{figure}[ht!]
\centering
\begin{subfigure}{.65\textwidth}
  \centering
  \includegraphics[width=1.0\linewidth]{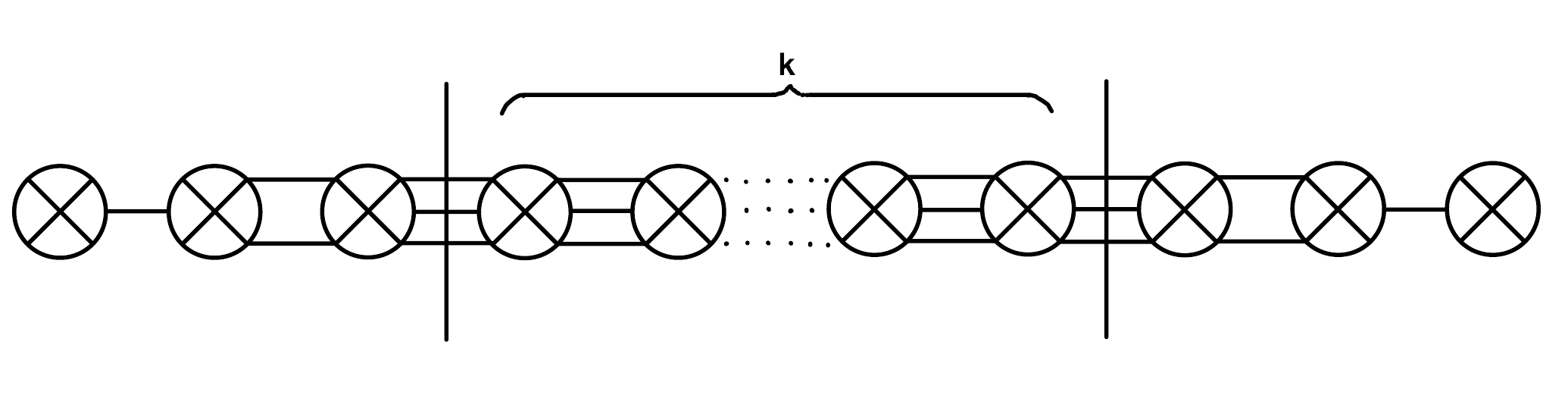}
  \caption{HW S-dual configuration.}
  \label{fig-sun_nf_mirror_HW}
\end{subfigure}%
\begin{subfigure}{.35\textwidth}
  \centering
  \includegraphics[width=0.9\linewidth]{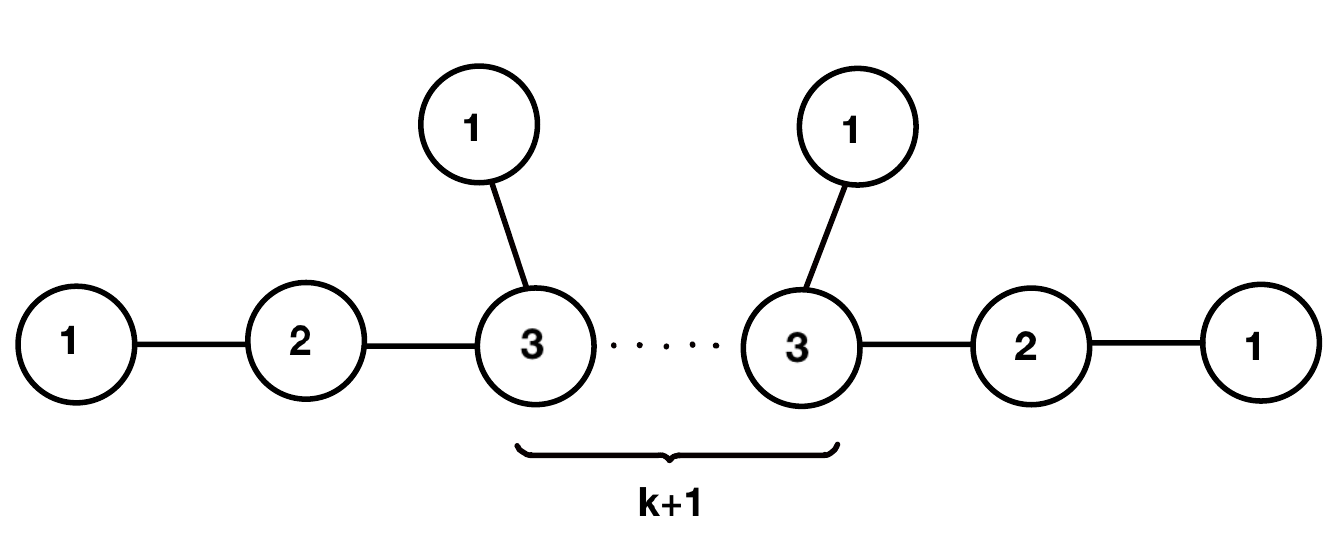}
  \caption{Corresponding quiver.}
  \label{fig-sun_nf_mirror_quiver}
\end{subfigure}
\caption{B-theory.}
\end{figure}

\begin{figure}[ht!]
\centering
\includegraphics[scale=.55]{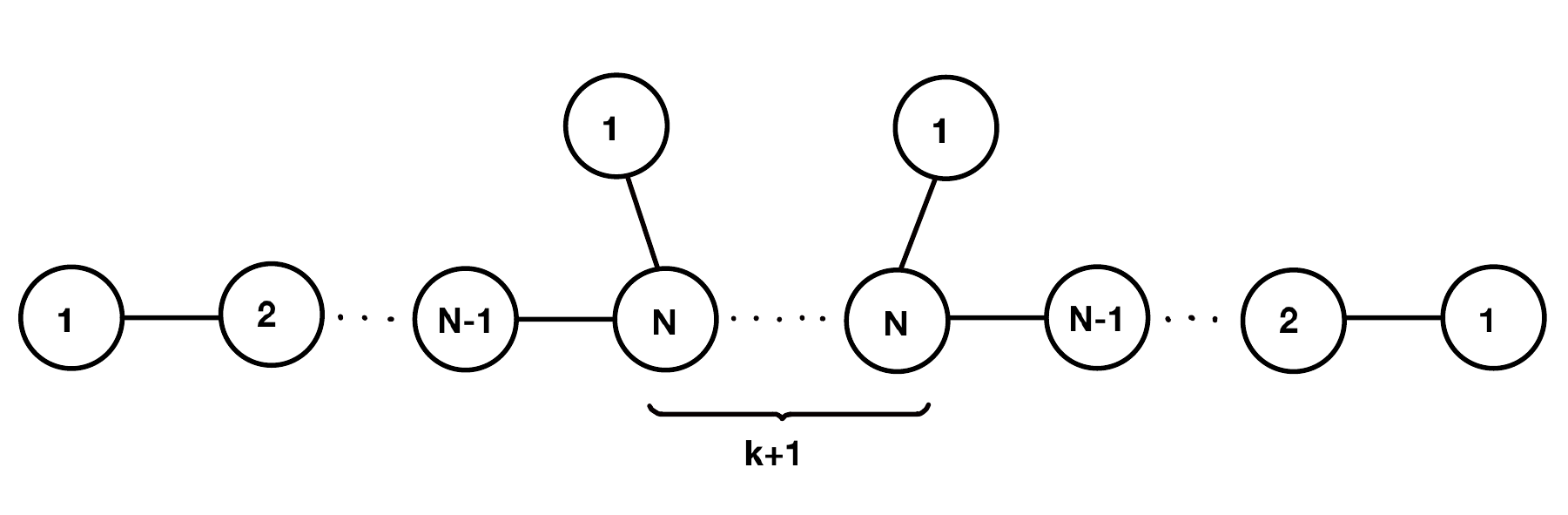}
\caption{Mirror quiver of $SU(N)$ with $N_f=2N+k$ flavors.}
\label{fig-crab}
\end{figure}

The S-dual configuration, shown in Figure~\ref{fig-sun_nf_mirror_HW}, gives rise to the quiver in Figure~\ref{fig-sun_nf_mirror_quiver}, which displays the B-theory. 
This can be generalized to $SU(N)$ with $2N+k$ flavors, whereby the B-theory is easily seen to yield the quiver in Figure~\ref{fig-crab}. This matches precisely the QFT predictions of the original \cite{Hanany:1996ie}. 

As an easy check, we should match the expected $U(N_f)$ flavor symmetry acting on the hypers of the original theory to the Coulomb branch symmetry of the mirror theory. This is easily confirmed by noticing that there is a balanced subquiver in the form of an $A_{N_f-1}$ Dynkin diagram, and that one of the two unbalanced $U(1)$ nodes\footnote{We can choose a basis where the other $U(1)$ is the decoupled one.} contributes the extra $U(1)_T$, thereby making a $U(N_f)$ symmetry. For the A-theory, we note that the Coulomb branch now has a trivial symmetry group, as does the Higgs branch of the B-theory. This is summarized in Table~\ref{tab:sun_nf_match}.
\begin{table}[th!]
\be
\begin{tabular}{c|c|c}
 & A-theory & B-theory \\ \hline
$G_{HB}$ & $U(N_f)$  & $\{ 1 \}$  \\
$G_{CB}$ & $\{ 1 \}$  & $U(N_f)$  \\
 \end{tabular}\nonumber
 \ee
\caption{Symmetry group of the Higgs Branch ($G_{HB}$) and of the Coulomb Branch ($G_{CB}$) for A-theory and B-theory.}
    \label{tab:sun_nf_match}
\end{table}

For the special case of $SU(2)$, we expect a $D_{N_f}$-symmetry on the Higgs branch of the A-side, and therefore a $D_{N_f}$-symmetry on the Coulomb branch of the B-side. Indeed, we see that we get a quiver in the shape of the extended $D_{N_f}$ Dynkin diagram. As usual, one gauge $U(1)$ decouples, so the symmetry is really just $D_{N_f}$.

\subsection{Homogeneous linear quivers}
\begin{figure}[h!]
\centering
\includegraphics[scale=.75]{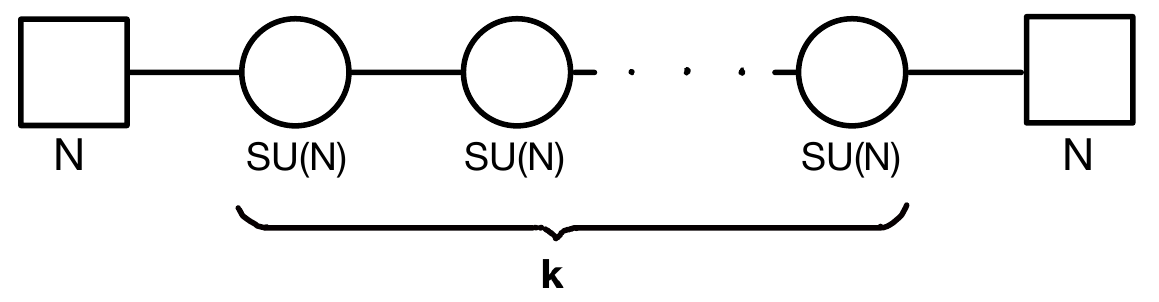}
\caption{Homogeneous linear quiver.}
\label{fig-homog_quiver}
\end{figure}

\begin{figure}[ht!]
\centering
\begin{subfigure}{.5\textwidth}
  \centering
  \includegraphics[width=1.0\linewidth]{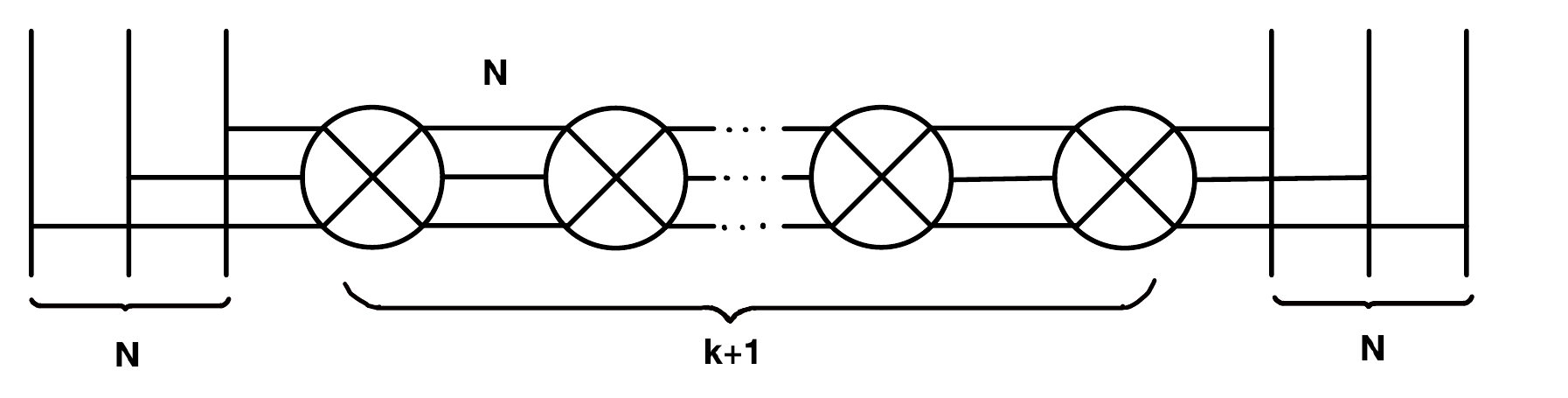}
  \caption{A-theory}
  \label{fig-homog_HW}
\end{subfigure}%
\begin{subfigure}{.5\textwidth}
  \centering
  \includegraphics[width=1.0\linewidth]{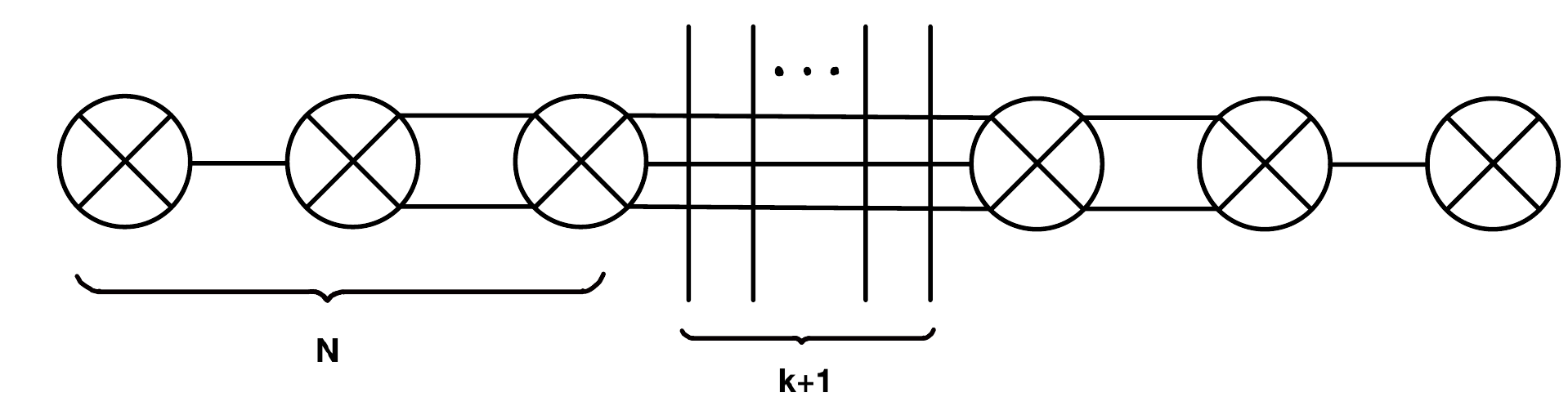}
  \caption{B-theory}
  \label{fig-homog_mirror_HW}
\end{subfigure}
\caption{HW setups of the homogeneous quiver $(\mathrm{a})$ and its mirror $(\mathrm{b})$.}
\end{figure}

\begin{figure}[h!]
\centering
\includegraphics[scale=.55]{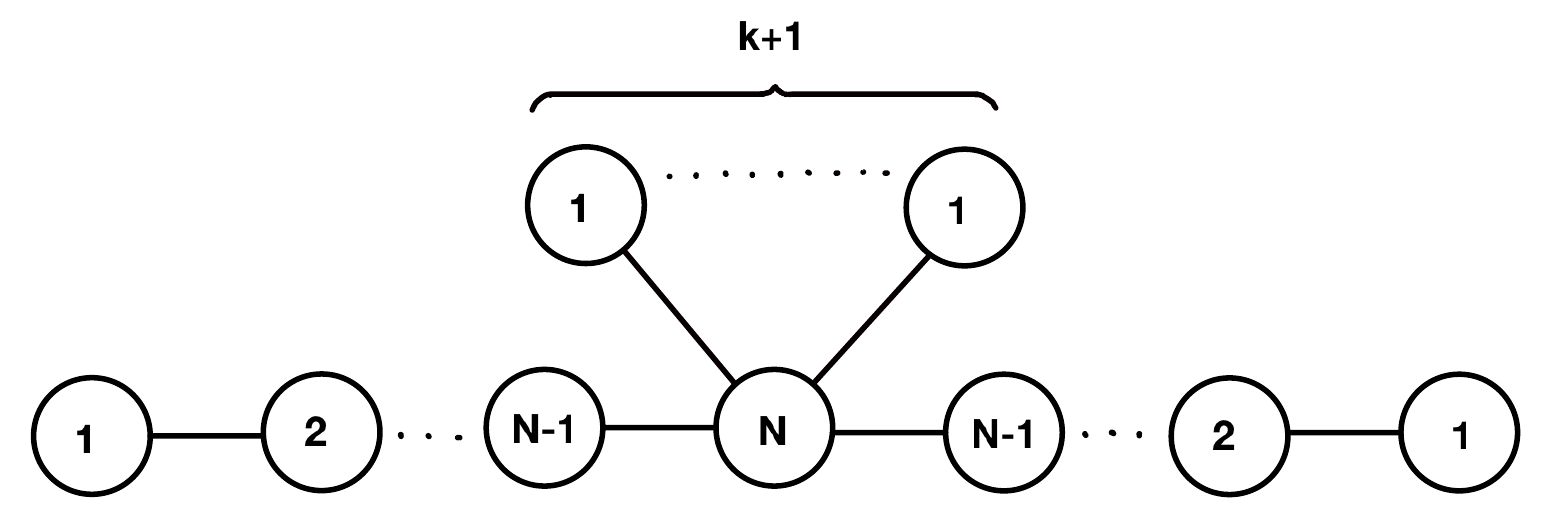}
\caption{Mirror dual of homogeneous linear quiver.}
\label{fig-homog_quiver_mirror}
\end{figure}

The previous section included as a subset the family $SU(N)$ with $2 N$ flavors. Let us generalize that to a fully balanced linear quiver of the form given in Figure~\ref{fig-homog_quiver}, with $k$ nodes. The HW setup is given in Figure~\ref{fig-homog_HW}, and its S-dual configuration in Figure~\ref{fig-homog_mirror_HW}, which yields the mirror quiver given by Figure~\ref{fig-homog_quiver_mirror}. 

On the A-side, there is a jump in the Higgs branch dimension by $k$ (i.e. the number of nodes). In our string theory setup, this is realized by the fact that $k+1$ NS5-branes can be freed to move along $x_{4, 5, 6}$, giving rise to $k$ relative positions.

The matching of the symmetries is interesting. On the A-side, the Coulomb branch symmetry becomes trivial. On the other hand, the Higgs branch symmetry is enhanced from $SU(N) \times SU(N)$ (corresponding to the two extremal flavor nodes), to $U(N) \times U(N) \times U(1)^{k-1}$, whereby the Abelian factors act as phases on the $k-1$ internal hypers.
On the B-side, we see that the Higgs branch has a trivial symmetry group, since all nodes are gauged. On the other hand, the Coulomb branch is interesting. There are two $A_{N-1}$ Dynkin subquivers to the left and to the right of the central node. Finally, the central node, and the $k+1$ nodes attached to it, provide a $U(1)^{k+1}$ topological symmetry (taking into account one decoupled $U(1)$).  In total, we see $SU(N) \times SU(N) \times U(1)^{k+1}$, which matches the Higgs branch symmetry of A-theory.

\subsection{Ascending linear quivers}
\begin{figure}[h!]
\centering
\includegraphics[scale=.65]{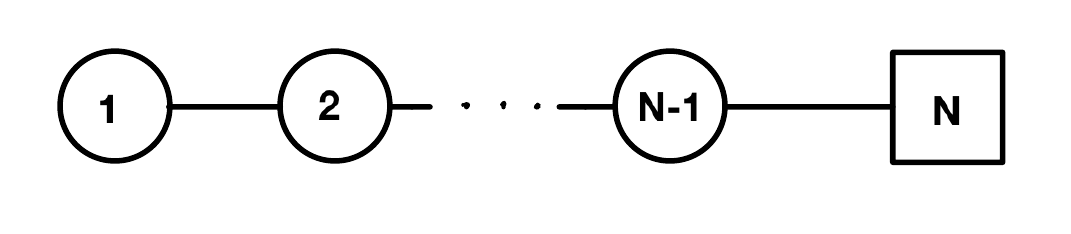}
\caption{$TSU(N)$ quiver.}
\label{fig-tsun}
\end{figure}
\begin{figure}[h!]
\centering
\includegraphics[scale=.75]{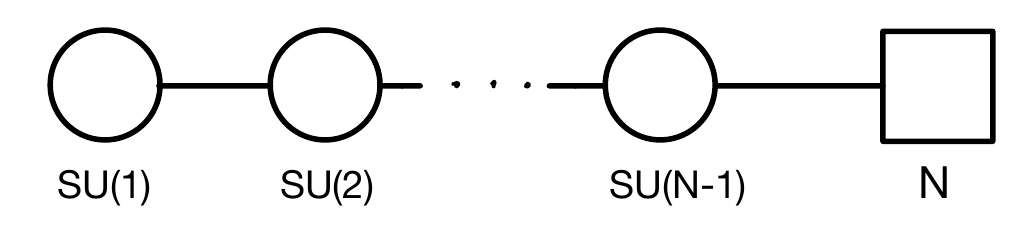}
\caption{A-theory quiver.}
\label{fig-STSUNquiver}
\end{figure}

Finally, we come to our last class of standard examples, the so-called $TSU(N)$ theories. This theory is defined by the quiver in Figure~\ref{fig-tsun}. It is known to be self-mirror. Its Higgs and Coulomb branch symmetry groups are $SU(N)$. After eliminating all $U(1)$'s from the gauge nodes, the quiver becomes the one in Figure~\ref{fig-STSUNquiver} and we expect the new Higgs branch symmetry to be $U(N) \times U(1)^{N-2}$. 

\begin{figure}[h!]
\centering
\includegraphics[scale=.65]{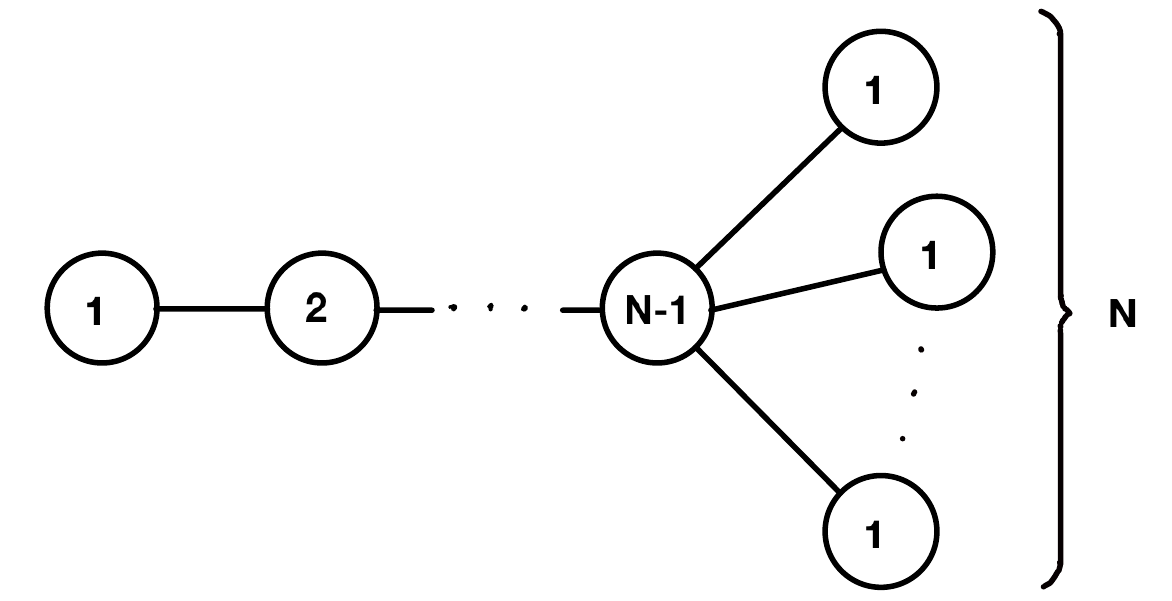}
\caption{B-theory quiver.}
\label{fig-stsun_mirrorquiver}
\end{figure}

For the mirror theory, the quiver is found analogously to the previous examples and is displayed in Figure~\ref{fig-stsun_mirrorquiver}. We see that the Coulomb branch symmetry becomes $SU(N) \times U(1)^{N-1}$, which matches the A-side.

\section{An outlier: $SU(2)$ with two flavors}
Now we come to an interesting example of a so-called `bad theory' in the sense of~\cite{Gaiotto:2008ak}. What makes such theories difficult to tackle, is the fact that they have monopole operators whose expected IR R-charge falls below the unitary bound. It is therefore not clear, whether they admit a conventional SCFT fixed point. In \cite{Assel:2017jgo} and \cite{Ferlito:2016grh}, such theories are studied from the field theory point of view. More precisely, their moduli spaces are constructed, and statements are made about mirror symmetry candidates. One difficulty arises, namely, the fact that their Higgs branches consist of unions of cones makes it unlikely to find a purely field-theoretic mirror dual.\footnote{We thank Julius Grimminger for explaining this to us.}

However, nothing precludes string theory from finding a mirror setup to these theories. Indeed, in this section, we will provide this, and note that those setups do not have conventional QFT descriptions. We will not comment on the moduli spaces, but rather focus on the construction itself.

\begin{figure}
\centering
\begin{subfigure}{.5\textwidth}
  \centering
  \includegraphics[width=1\linewidth]{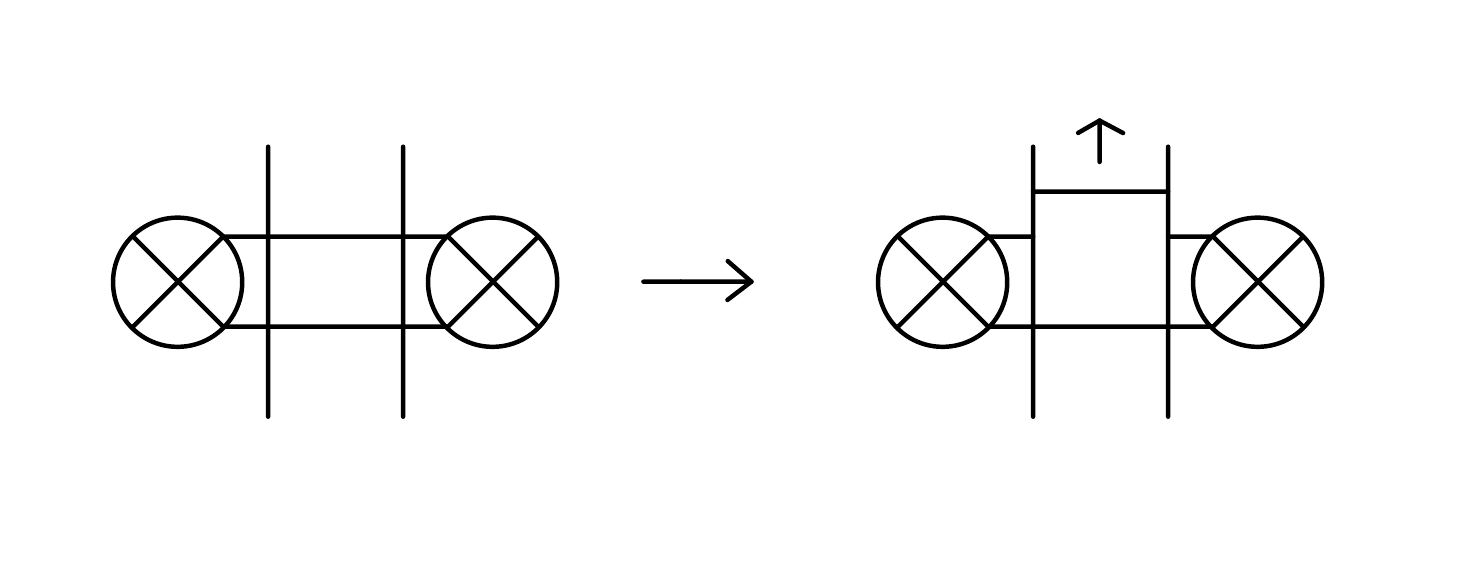}
  \caption{`Mesonic' branch.}
  \label{fig:su2_2_meson}
\end{subfigure}%
\begin{subfigure}{.5\textwidth}
  \centering
  \includegraphics[width=1\linewidth]{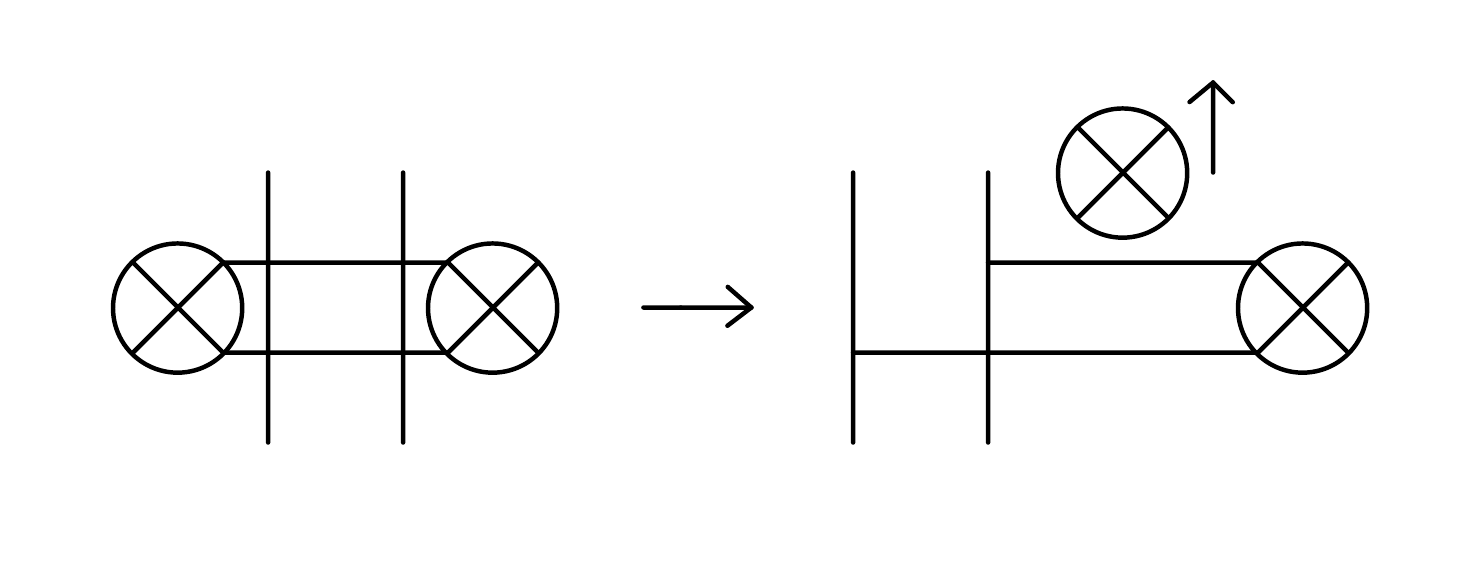}
  \caption{`Baryonic' branch.}
  \label{fig:su2_2_baryon}
\end{subfigure}
\caption{$SU(2)$ with two flavors.}
\label{fig:test}
\end{figure}

A simple case in point is $SU(2)$ with two doublets, which was studied in \cite{Assel:2018exy}. Its Higgs branch consists of the union of two copies of $\cc^2/\zz_2$. Our IIB setup for this theory is very simple: Two D3's suspended between two NS5's, with two D5's intersecting the D3's, and, as usual, we compactify the transverse NS5 directions. Can we see the two cones comprising the Higgs branch? In Figure~\ref{fig:su2_2_meson} we show the `mesonic' cone. This is the piece that would have been present also for $U(2)$ with two flavors.
Figure~\ref{fig:su2_2_baryon} shows the new `baryonic' branch, which requires one NS5 to be released. Clearly, these two branches are mutually exclusive, as expected for a union of cones. 

The first cone is clearly isomorphic to $\cc^2/\zz_2$. In order to see this, we can first move the lower D3-segment away from the page along the NS5-branes. Now, locally, we have the brane situation of a $U(1)$ with two hypers, whose Higgs branch is known to be the $A_1$-singularity. 

The new `baryonic' branch is more subtle. We see that it is quaternionic one-dimensional, since there is just one mobile NS5. However, we can also see that this space is an $A_1$-singularity as follows: The mobile NS5 hosts a $U(1)$ vector multiplet. The D3-segments are suspended between an NS5 and one D5 each, and therefore host nothing. On the other hand, we can have D1-strings suspended between the D3's and the mobile NS5. We recognize this theory as simply SQED$_2$, which has $\cc^2/\zz_2$ as its Higgs branch.

\begin{figure}[ht!]
\centering
\includegraphics[scale=.6]{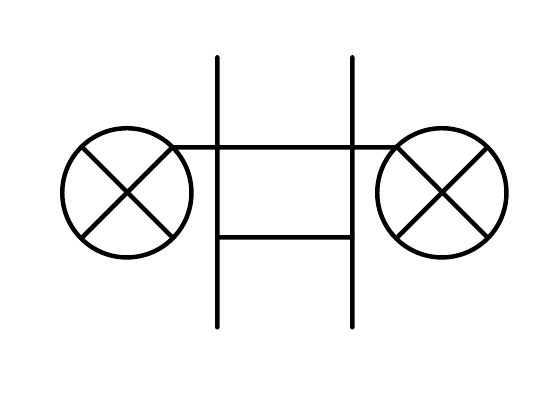}
\caption{Mirror dual of $SU(2)$ with two flavors.}
\label{fig:unconv}
\end{figure}

Having defined this setup, and demonstrated that it has the expected Higgs branch structure, let us now draw the mirror symmetry setup in IIB. What becomes immediately clear, is that the Hanany-Witten dual admits no field theory interpretation, as there is no way to appropriately disentangle the brane segments into a configuration where a low-energy theory can be read off. In Figure~\ref{fig:unconv}, we show what this looks like.
Now, by definition, this setup has a `Coulomb branch' with the same structure as the Higgs branch of the original theory, even though no Lagrangian definition will reproduce it.

\section*{Acknowledgments} 

We have benefited from discussions with Sergio Benvenuti, Antoine Bourget and Simone Giacomelli. A.C.~is a Research Associate of the Fonds de la Recherche Scientifique F.N.R.S.~(Belgium). The work of A.C.~is partially supported by IISN - Belgium (convention 4.4503.15), and supported by the Fonds de la Recherche Scientifique - F.N.R.S.~under Grant CDR J.0181.18. 
The work of R.V.~is partially supported by ``Fondo per la Ricerca di Ateneo - FRA 2018'' (UniTS) and by INFN Iniziativa Specifica ST\&FI.





\begin{thebibliography}{10}

\bibitem{Hanany:1996ie}
A.~Hanany and E.~Witten, ``{Type IIB superstrings, BPS monopoles, and
  three-dimensional gauge dynamics},'' {\em Nucl. Phys. B} {\bf 492} (1997)
  152--190, \href{http://arXiv.org/abs/hep-th/9611230}{{\tt hep-th/9611230}}.

\bibitem{Intriligator:1996ex}
K.~A. Intriligator and N.~Seiberg, ``{Mirror symmetry in three-dimensional
  gauge theories},'' {\em Phys. Lett. B} {\bf 387} (1996) 513--519,
  \href{http://arXiv.org/abs/hep-th/9607207}{{\tt hep-th/9607207}}.

\bibitem{deBoer:1996mp}
J.~de~Boer, K.~Hori, H.~Ooguri, and Y.~Oz, ``{Mirror symmetry in
  three-dimensional gauge theories, quivers and D-branes},'' {\em Nucl. Phys.
  B} {\bf 493} (1997) 101--147, \href{http://arXiv.org/abs/hep-th/9611063}{{\tt
  hep-th/9611063}}.

\bibitem{Porrati:1996xi}
M.~Porrati and A.~Zaffaroni, ``{M theory origin of mirror symmetry in
  three-dimensional gauge theories},'' {\em Nucl. Phys. B} {\bf 490} (1997)
  107--120, \href{http://arXiv.org/abs/hep-th/9611201}{{\tt hep-th/9611201}}.

\bibitem{Witten:2003ya}
E.~Witten, ``{SL(2,Z) action on three-dimensional conformal field theories with
  Abelian symmetry},'' in {\em {From Fields to Strings: Circumnavigating
  Theoretical Physics: A Conference in Tribute to Ian Kogan}}, pp.~1173--1200.
\newblock 7, 2003.
\newblock \href{http://arXiv.org/abs/hep-th/0307041}{{\tt hep-th/0307041}}.

\bibitem{Dey:2014tka}
A.~Dey, A.~Hanany, P.~Koroteev, and N.~Mekareeya, ``{Mirror Symmetry in Three
  Dimensions via Gauged Linear Quivers},'' {\em JHEP} {\bf 06} (2014) 059,
  \href{http://arXiv.org/abs/1402.0016}{{\tt 1402.0016}}.

\bibitem{Dey:2020hfe}
A.~Dey, ``{Three Dimensional Mirror Symmetry beyond $ADE$ quivers and
  Argyres-Douglas theories},'' \href{http://arXiv.org/abs/2004.09738}{{\tt
  2004.09738}}.

\bibitem{Klebanov:2010tj}
I.~R. Klebanov, S.~S. Pufu, and T.~Tesileanu, ``{Membranes with Topological
  Charge and AdS${_4}$/CFT$_{3}$ Correspondence},'' {\em Phys. Rev. D} {\bf 81}
  (2010) 125011, \href{http://arXiv.org/abs/1004.0413}{{\tt 1004.0413}}.

\bibitem{Benishti:2010jn}
N.~Benishti, D.~Rodriguez-Gomez, and J.~Sparks, ``{Baryonic symmetries and M5
  branes in the $AdS_{4}/CFT_{3}$ correspondence},'' {\em JHEP} {\bf 07} (2010)
  024, \href{http://arXiv.org/abs/1004.2045}{{\tt 1004.2045}}.

\bibitem{Gaiotto:2008ak}
D.~Gaiotto and E.~Witten, ``{S-Duality of Boundary Conditions In N=4 Super
  Yang-Mills Theory},'' {\em Adv. Theor. Math. Phys.} {\bf 13} (2009), no.~3,
  721--896, \href{http://arXiv.org/abs/0807.3720}{{\tt 0807.3720}}.

\bibitem{Assel:2017jgo}
B.~Assel and S.~Cremonesi, ``{The Infrared Physics of Bad Theories},'' {\em
  SciPost Phys.} {\bf 3} (2017), no.~3, 024,
  \href{http://arXiv.org/abs/1707.03403}{{\tt 1707.03403}}.

\bibitem{Hanany:1997gh}
A.~Hanany and A.~Zaffaroni, ``{Branes and six-dimensional supersymmetric
  theories},'' {\em Nucl. Phys. B} {\bf 529} (1998) 180--206,
  \href{http://arXiv.org/abs/hep-th/9712145}{{\tt hep-th/9712145}}.

\bibitem{Brunner:1997gf}
I.~Brunner and A.~Karch, ``{Branes at orbifolds versus Hanany Witten in
  six-dimensions},'' {\em JHEP} {\bf 03} (1998) 003,
  \href{http://arXiv.org/abs/hep-th/9712143}{{\tt hep-th/9712143}}.

\bibitem{Grimm:2011tb}
T.~W. Grimm, M.~Kerstan, E.~Palti, and T.~Weigand, ``{Massive Abelian Gauge
  Symmetries and Fluxes in F-theory},'' {\em JHEP} {\bf 12} (2011) 004,
  \href{http://arXiv.org/abs/1107.3842}{{\tt 1107.3842}}.

\bibitem{Collinucci:2020jqd}
A.~Collinucci and R.~Valandro, ``{The role of U(1)'s in 5d theories, Higgs
  branches, and geometry},'' \href{http://arXiv.org/abs/2006.15464}{{\tt
  2006.15464}}.

\bibitem{Kapustin:1999ha}
A.~Kapustin and M.~J. Strassler, ``{On mirror symmetry in three-dimensional
  Abelian gauge theories},'' {\em JHEP} {\bf 04} (1999) 021,
  \href{http://arXiv.org/abs/hep-th/9902033}{{\tt hep-th/9902033}}.

\bibitem{Ferlito:2016grh}
G.~Ferlito and A.~Hanany, ``{A tale of two cones: the Higgs Branch of Sp(n)
  theories with 2n flavours},'' \href{http://arXiv.org/abs/1609.06724}{{\tt
  1609.06724}}.

\bibitem{Assel:2018exy}
B.~Assel and S.~Cremonesi, ``{The Infrared Fixed Points of 3d $\mathcal{N}=4$
  $USp(2N)$ SQCD Theories},'' {\em SciPost Phys.} {\bf 5} (2018), no.~2, 015,
  \href{http://arXiv.org/abs/1802.04285}{{\tt 1802.04285}}.

\end{thebibliography}

\providecommand{\href}[2]{#2}

\end{document}